\pdfoutput=1
\documentclass[traditabstract]{aa}  
\usepackage{graphicx}
\usepackage{txfonts}
\begin{document}
\title{A critical evaluation of PCA detection of polarized signatures
  using real stellar data}

   \author{
          Fr\'ed\'eric Paletou
          }

          \institute{Universit\'e de Toulouse, UPS-Observatoire
            Midi-Pyr\'en\'ees, Irap, Toulouse, France\\
\and
CNRS,  Institut de Recherche en Astrophysique et
            Plan\'etologie, 14 av. E. Belin, 31400 Toulouse, France\\
            \email{fpaletou@irap.omp.eu}
}

   \date{Received April 12, 2012; accepted June 22, 2012}


   \abstract 
   {The general context of this study concerns the post-processing of
     multiline spectropolarimetric observations of stars, and in
     particular these numerical analysis techniques aiming at the
     detection and the characterization of polarized signatures.
  Hereafter, using real observational data, we compare
    and clarify a number of points concerning various
    methods of analysis.
    Indeed, simple line addition, least-squares deconvolution and
    denoising by principal component analysis have been applied, and
    compared to each other, to polarized stellar spectra available
    from the TBLegacy database of the Narval spectropolarimeter.
    Such a comparison between various approaches of distinct
    sophistication levels allows us to make a safe choice for the next
    implementation of on-line post-processing of our unique
    database for the stellar physics community. }
   {}


   \maketitle

%

\section{Introduction}

The present study concerns the post-processing of multiline
spectropolarimetric measurements, and in particular of stellar
data. We focus hereafter on data collected, since 2006, with the Narval
spectropolarimeter mounted at the 2-m aperture TBL telescope located
at the summit of the \emph{Pic du Midi de Bigorre} (France). We
investigate, in particular, the capabilities of principal component
analysis (hereafter PCA) on observations made with Narval.

PCA has been regularly used in \emph{solar} spectropolarimetry during the
last decade (see e.g., Rees et al. 2000 and Skumanich \& L\'opez Ariste
2002). Its main purpose was to provide an alternative way of inverting
spectropolarimetric data, for the determination of the vector magnetic
field present in various solar features, from sunspots to solar
prominences (see e.g., L\'opez Ariste \& Casini 2002).

Concerning stellar data, PCA-based denoising of spectral lines was
first presented by Caroll et al. (2007). It was further tested on data
taken with the SOFIN spectrograph at the NOT telescope. This procedure
was mainly driven by the purpose of doing Zeeman-Doppler Imaging
(hereafter ZDI; see Semel 1989) from temporal sequences of individual
spectral lines, instead of using pseudo-profiles such as the ones
commonly computed by least-squares deconvolution (hereafter LSD; see
Donati et al. 1997 and Kochukhov et al. 2010, for a recent
review and discussion).  More recently, Mart\'{i}nez Gonz\'alez et
al. (2008) discussed in details the capabilities of PCA denoising of
solar and stellar spectropolarimetric data, using \emph{synthetic}
data. They also provided some comments concerning the relationship
between PCA denoising, line addition and least-squares
deconvolution. Later on, Ram\'{i}rez V\'elez et al. (2010) proposed
another PCA-based method, coupled to ZDI, which was applied to a very
limited set of observational data taken both at the AAT telescope with
the SemelPol spectropolarimeter, and with Narval at the TBL.

Hereafter we come back on some details of PCA denoising and analysis
of \emph{observational} spectropolarimetric data. We discuss further
the practical capabilities of such an approach. Comparisons with LSD
and the so-called (simple) line addition (hereafter SLA; Semel et
al. 2009) methods are also discussed.


\section{The source of data}

We have been using Narval data available from the \emph{public}
database TBLegacy\footnote{http://tblegacy.bagn.obs-mip.fr/}. Narval
is a state-of-the-art spectropolarimeter operating in the 0.38-1
$\mu$m spectral domain, with a spectral resolution of 65\,000 in
  its polarimetric mode. It is an improved copy, adapted to the
  2-m TBL telescope, of the Espadons spectropolarimeter, in
operations since 2004 at the 3.6-m aperture CFHT telescope (see Donati
et al. 2006 for further technical details).

The TBLegacy database is operational since 2007. It is at the
  present time the largest on-line archive of high-resolution
  polarization spectra. It hosts data which were taken at the 2-m
TBL telescope since december 2006. So far, more than 70\,000 spectra
have been made available, for more than 370 distinct targets all over
the Hertzsprung-Russell diagram. More than 13\,000 \emph{polarized}
spectra are also available, mostly for \emph{circular} polarization
(linear polarization data are very seldom still and amounts to a few
hundreds spectra, but it is equally available). By default, the latter
is the usual circular polarization $V(\lambda) /I_c$ normalized to the
local continuum intensity.

At the present time, the TBLegacy database provides no more than
Stokes $I$ or $V /I_c$ spectra calibrated in wavelength.  Stokes $I$ data
are either normalised to the local continuum or not. In a next step,
further post-processing of these spectra will be proposed on-line to
users and the relevant software will be made fully available to the
community. It will be the case for the simple line addition and the
least-squares deconvolution standard procedures that we shall be using
in the present study, together with PCA denoising.


\begin{figure}
  \centering
  \includegraphics[width=9cm,angle=0]{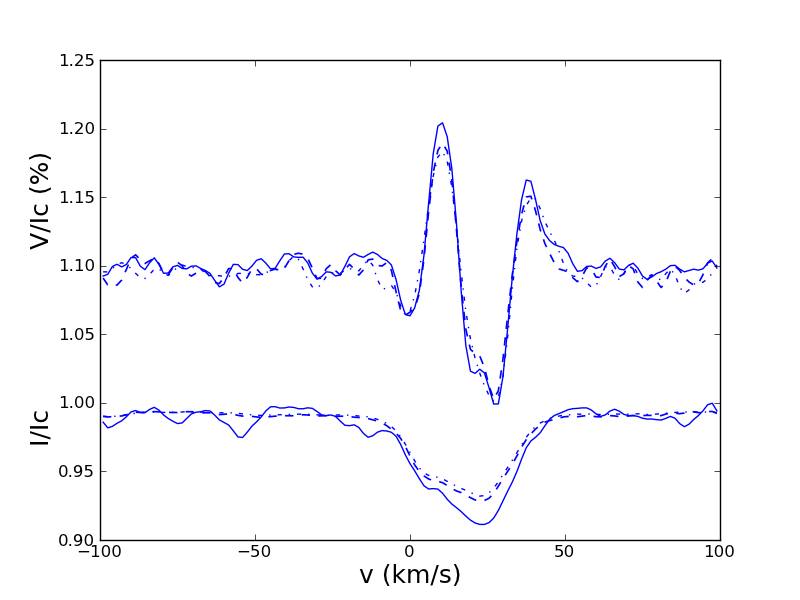}
  \caption{Comparison between LSD (full lines) and SLA (dashed lines)
    $I/I_{\rm c}$ and $V/I_{\rm c}$ pseudo-line profiles, for II Peg
    observations of August 2008. Stokes V profiles have been shifted
    by 1.1 so the largest amplitude lobe, for LSD, is about 0.1\% of
    $I_{\rm c}$ in that case. $P_1$ profiles (dot-dashed) both for
    $I/I_{\rm c}$ and $V/I_{\rm c}$ resulting from PCA analysis of the
    data are also displayed for comparison purpose.}
  \label{Fig1}
\end{figure}

\section{Numerical procedures}

\subsection{The matrix of observations}

Observations we get from TBLegacy are basically Stokes $I(\lambda)$ or
$V(\lambda)$. Each of them consist in a very large array of about
  200\,000 elements covering the whole spectral domain observable by
  Narval. The main task of building the matrix of observations
$\vec{O}$ is to split the multiline observations vs. wavelength into
$N_{\rm obs}$ elementary profiles, each of them centered at a
  given wavelength and projected onto a common \emph{velocity}
  grid. Such a velocity grid is an \emph{a priori} data that we adopt
in our numerical procedure. The choice of such a grid of velocities
depends on the spectral sampling of the original set of data --
in the case of Narval data it is of the order of 1.8 ${\rm km}.{\rm
  s}^{-1}$, as well as on the target nature, for what concerns the
velocity range to be considered (typically between $\pm 120$ and $\pm
200$ ${\rm km}.{\rm s}^{-1}$). Practically we shall be dealing with
$N_v$ velocity bins of the order of $10^2$, while $I$ or $V$ original
data (sampled in wavelength), will be recast into $N_{\rm obs}$
  elementary $v$-sampled profiles, where $N_{\rm obs}$ is of the order
  of $10^3$-$10^4$, depending on the spectral type of the target.

\begin{figure}
  \centering
  \includegraphics[width=9cm,angle=0]{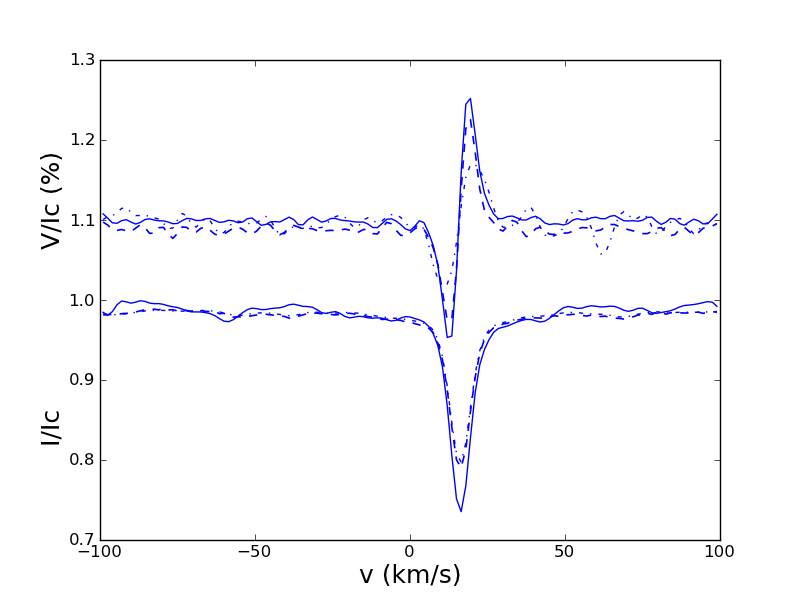}
  \caption{Same as Fig. (1) but for $\varepsilon$ Eri observations of
    February 2007. Stokes V profiles have been shifted by 1.1 and
    multiplied by 3, so the largest amplitude lobe, for LSD, is about
    0.05\% of $I_{\rm c}$ in that case.}
  \label{Fig2}
\end{figure}

The transformation of the original data requires the help of a
supplementary file, usually called ``mask'' and which consists in the
list of all the wavelengths at rest, $\lambda_0$, of the spectral
lines expected to be present in the observations of a given spectral
type of stars. For all of the cases discussed hereafter, we have used
mask files widely used by the community and built from the VALD
database (Piskunov et al. 1995; ressources for this study have been
kindly provided to us by E. Alecian). In general, these mask
  files contain additional information about each spectral line, in
  particular their line depression, $d_i$, and effective Land\'e
  factors, $g_i$, required by LSD (see next section).

Therefore, given a proper mask and a velocity grid, it is quite easy
and straightforward to transform $I(\lambda)$ or $V(\lambda)$ data
into $N_{\rm obs}$ individual $I(v)$ or $V(v)$ profiles, in accordance
with the Doppler--Fizeau effect and the well-known relationship

\begin{equation}
{{\delta v} \over {c}}= {{\delta \lambda} \over {\lambda_0}}  \, ,
\end{equation} 
where $\delta \lambda = (\lambda- \lambda_0)$, which is computed from
the original data (see also \S2.1 in Ram\'{i}rez V\'elez et al. 2010).

This operation results into the construction of a ($N_{\rm obs}$,$N_v$)
rectangular matrix of observations $\vec{O}$ which shall now be used in
different ways.

\subsection{Simple line addition vs. LSD}

For a wealth of data in TBLegacy, down to polarized signatures $V/I_c$
of the order of 0.01\%, the pseudo-profiles resulting from the simple
line addition (or, to be more precise the \emph{unweighted}, or
arithmetic mean) of the $N_{\rm obs}$ individual spectral lines of
$\vec{O}$ are very meaningful, both from the standpoints of the
detection and of the characterization (i.e., the proper determination
of its shape and amplitudes) of the polarized signature carried by the
multiline, but noisy, observations. Moreover, SLA profiles are very
similar to the one obtained from least-squares deconvolution. This was
indeed mentioned and discussed in the very instructive, but
unfortunately overlooked at, recent article of Semel et
al. (2009). Nevertheless, to the best of our knowledge, \emph{no
  direct comparisons between LSD and SLA profiles obtained with real
  data such as Narval's, have been published yet}.

To remedy that, in Figs. (1) and (2), we display both LSD and SLA
pseudo-profiles obtained directly by computing a \emph{simple average}
of all the rows of the $\vec{O}$ matrix constructed from the same set
of observations of the RS CVn star II Peg, made in August 2008.  The
LSD profiles for Stokes $I$ have been computed using weights $\omega_I
= d_i$ normalized to the arithmetic mean of the considered central
line depressions $d_i$. Those for Stokes $V$ were computed for weights
$\omega_V = g_i \lambda_{0, i} d_i$ normalized to the arithmetic mean
of the $\omega_V$'s. Also, \emph{no} line depth cut-off criterium was
adopted there (provided that depressions are, originally, greater
  than or equal to 10\% of the continuum). Considerations and
recommendations about the issue of LSD weights definition (and
especially their normalisation) and line depth cut-off criterium can
be found in Kochukhov et al. (2010). The latter revealed some
indiscipline in the community of LSD users and subsequent articles
still fail, unfortunately, in providing details about the exact
procedure which was applied to data -- see e.g., Kochukhov et
al. (2011) or Donati et al. (2011). To conclude on these points, we
again recommend this community to read carefully Semel et al. (2009)
and, especially, their \S 2.3 dedicated to the statistical
  properties of (LSD) weights.

For II Peg, we considered about 6600 wavelengths in the mask, covering
a 400-1000 nm range, using VALD data for a $T_{\rm eff.}  = 5\,000$ K,
a surface gravity of logg=3.0 cgs and solar abundances. Concerning
this choice of stellar parameters, Berdyugina et al. (1998) determined
$T_{\rm eff.} = 4\,600$ K and logg=3.2 cgs. However $T_{\rm eff.}$ as
high as 5\,250 K are still reported by VizieR. As can be seen in
Fig.\,(1), respective shapes of $I/I_{\rm c}$ and $V/I_{\rm c}$ are
well recovered, both by LSD and SLA, and they are indeed very
similar. Amplitudes of $I/I_{\rm c}$ and $V/I_{\rm c}$ LSD
pseudo-profiles appear systematically slightly larger than SLA
ones. However, this is not going to impair significantly any further
determination of the mean line-of-sight magnetic field usually made,
assuming the weak-field regime of the Zeeman effect, using the centre
of gravity method (Rees \& Semel 1979, and references therein).

We noticed similar effects, displayed in Fig.\,(2), using observations
of the K2V star $\varepsilon$ Eri made on February 2007 with Narval. For
that case, and after inspection of all VizieR ressources, we adopted a
$T_{\rm eff.}  = 5\,000$ K and a surface gravity of logg=4.5 cgs (and
solar abundances) mask (see also Koleva \& Vazdekis 2012).<

Using a test version of TBLegacy currently under
development\footnote{The (Python) software implemented for such an
  analysis will be made public, although it is already available upon
  request to the author.}, we have been able to verify indeed how
similar SLA and LSD signatures are, from the analysis of many other
cases including for hotter magnetic stars than the ones discussed in
this article.

\begin{figure}
  \centering
  \includegraphics[width=9cm,angle=0]{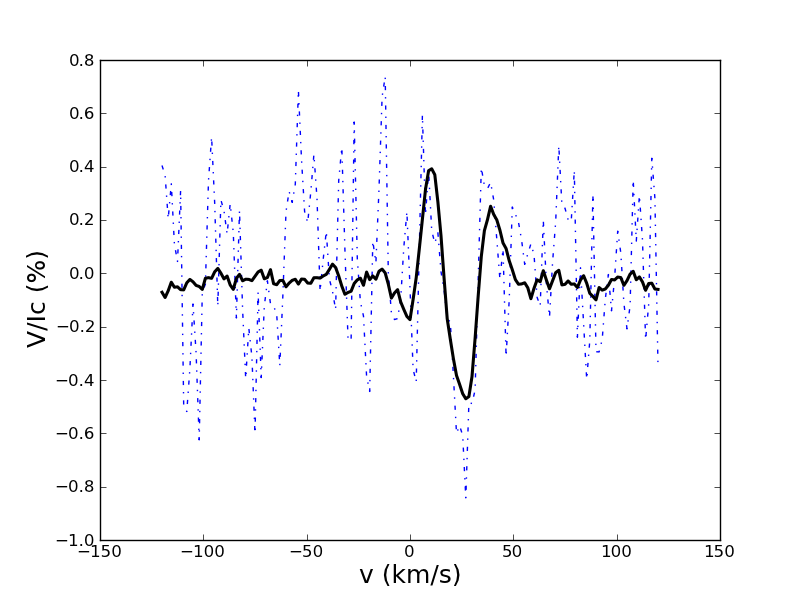}
  \caption{Example of PCA denoising: the original noisy signal
    $\vec{O}_j(v)$ (dashed line) for the 612.2 nm line of Ca\,{\sc i},
    is displayed, together with its projection on the eigenprofile of
    matrix $\vec{C}$ associated with the largest eigenvalue,
    $\vec{P}_{j,1}$ (full thick line). The latter profile already
    bears a shape very similar to the SLA (or LSD) pseudo-profiles
    obtained with the \emph{whole} set of observations, and displayed
    in Fig. (1).}
  \label{Fig3}
\end{figure}

\subsection{Principal component analysis}

Following Mart\'{i}nez Gonz\'alez et al. (2008), we built the
cross-product matrix, $\vec{C}=\vec{O}^T \vec{O}$, and computed its
eigenvalues $s_i$ and eigenvectors $\vec{e}_i$ (hereafter,
eigenprofiles). Hereafter we shall call $\vec{O}_j(v)$ the observation
made at wavelength index $j$, and we shall omit the dependance in $v$
of each of these individual profiles. \emph{No} physical assumption
  about the line formation process or the origin of the polarization
  signals are required for the PCA analysis we have carried-out. 

As demonstrated by Mart\'{i}nez Gonz\'alez et al. (2008) with their
Figs.~(1), without any noise (or a limited amount of it -- this is the
case for Stokes $I$ data from TBLegacy, for instance), the examination
of the sequence of eigenvalues $s_i$ of $\vec{C}$ shows that a few of
them will dominate, sometimes by orders of magnitudes as compared to
the smallest ones. However, for significant noise levels, as it will
be the case hereafter for Stokes $V$ data, the sequence of $s_i$ is in
general very slowly decreasing - see e.g., Fig.\,(6).

\begin{figure}
  \centering
  \includegraphics[width=9cm,angle=0]{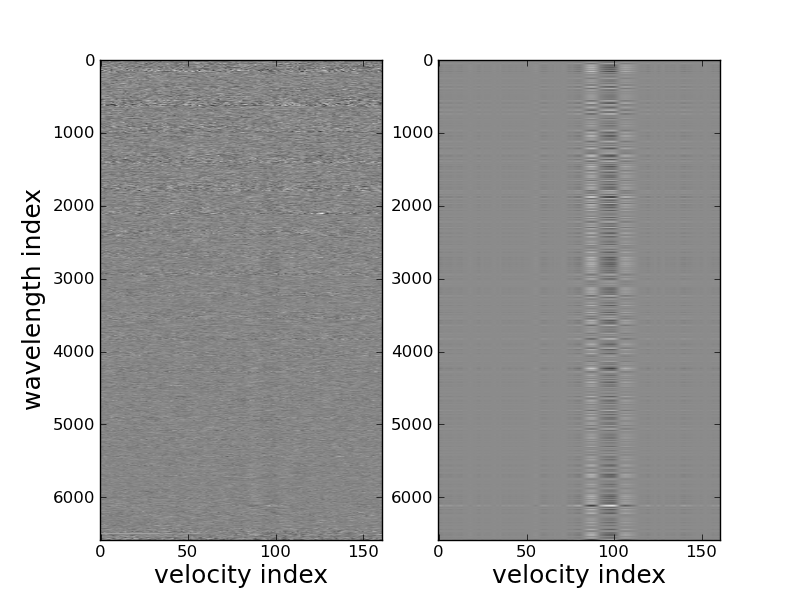}
  \caption{Comparison between the original map of the matrix of
    observations $\vec{O}$ (left) and the map of the $P_{j,1}$
    (right), for II Peg observations of August 2008. The efficiency of
    the PCA denoising is obvious at almost all wavelengths.}
  \label{Fig4}
\end{figure}

Even though the sequence of eigenvalues $s_i$ is very slowly decreasing
for most of Stokes $V$ data from TBLegacy, we first tried PCA
denoising by computing

\begin{equation}
\vec{P}_{j,k}= (\vec{O}_j \cdot \vec{e}_k) \vec{e}_k  \,
\end{equation} 
for $k$=1. For the wavelength index $j$ corresponding to the
  strong magnetically sensitive 612.2 nm line of Ca\,{\sc i} ,
Fig. (3) shows the efficiency of PCA denoising using only the
projection onto the eigenprofile $\vec{e}_1$ associated to the largest
eigenvalue $s_1$.  In that case, it is quite obvious that, for that
level of signal-to-noise ratio the gain provided by the PCA denoising
procedure is significant enough, and potentially allows for the
\emph{detection} of a meaningful polarized signature buried into
noise. Moreover, it is easy to notice that the single $\vec{P}_{j,1}$
denoised profile displayed in Fig. (3) already bears a shape very
similar to the SLA (or LSD) pseudo-profiles obtained from \emph{the
  whole set} of observations, as displayed in Fig. (1).

The efficiency of PCA denoising \emph{at all wavelengths} can also be
seen in Fig. (4) where we displayed images of the observations matrix
$\vec{O}$ (left) in comparison with the matrix of the
$\vec{P}_{j,1}$'s (right). In that case, clear polarized signatures
emerge almost at all observed wavelengths. It also opens the
possibility of a direct exploitation of single line data, instead of a
pseudo-profile combining all of the multiline signatures. The same is
true for $\varepsilon$ Eri data, for instance, even though its SLA (or
LSD) signature is significantly smaller than II Peg's.

\section{Comparison with SLA and LSD}

We have shown with the previous examples how PCA denoising can be
efficient on real stellar data. It can be very useful for
\emph{detection} purpose but \emph{could it offer an alternative to
  LSD or SLA methods?}

The case of $\varepsilon$ Eri is interesting in the sense that its
polarization signature is less complex but of much less amplitude than
the one of II Peg. Both LSD and SLA pseudo-line profiles, that is:

\begin{equation}
\bar{O}={1 \over {N_{\rm obs}}}{\sum_ {j=1}^{N_{\rm obs} }}
{\vec{O}_{j} (v)} \,
\end{equation}
show a clear antisymmetric $V/I_c$ profile with amplitudes of both
negative and positive lobe about 0.04-0.05\%, and spanning over a $
\Delta v \approx 30\, {\rm km}.{\rm s}^{-1}$ spectral range. But this
signature, recovered with two distinct methods, is not fully recovered
when we consider just the mean of the projection of the $\vec{O}_j$'s
onto eigenprofile $e_1$ only. The resulting mean profile is still
about a factor of 2 less the amplitude of $\bar{O}$ and the lobes are
also wider than the ones of LSD or SLA pseudo-profiles -- see again
Fig.\,(2).  Beyond the detection capability of PCA-based denoising,
this opens the further question of the proper characterization of the
``most common'' polarization signal content of the multiline
observations.

\begin{figure}
  \centering
  \includegraphics[width=9cm,angle=0]{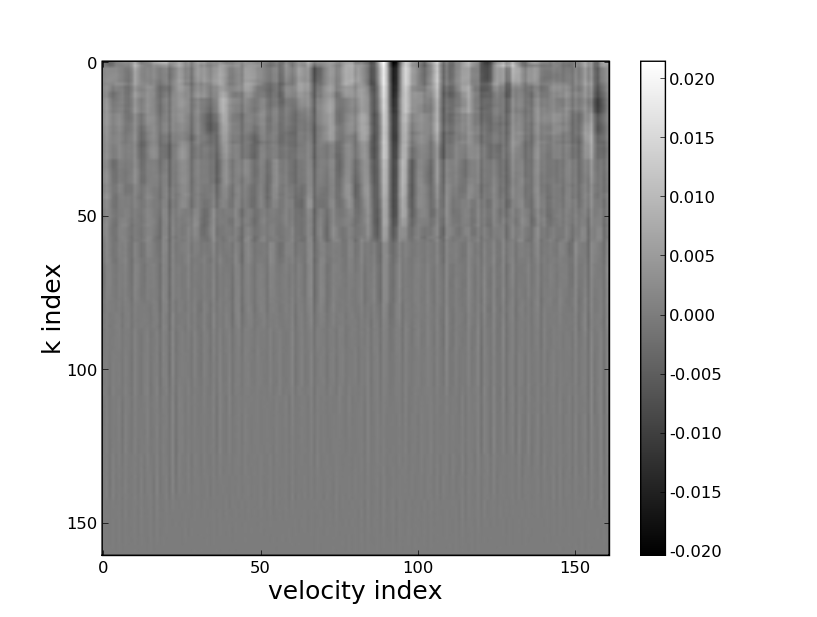}
  \caption{Map of the ($P_k - \bar{O}$) from circular polarization
    observations of $\varepsilon$ Eri made on February 2007. It takes
    about $k$=50-60 eigenprofiles for recovering the mean $\bar{O}$
    profile. The color scale on the right side of the image also
    indicates, in that case, that $P_{k=1}$ can be a factor of 2
    smaller in amplitude than the SLA mean profile whose amplitude is
    about 0.04\%.}
  \label{Fig5}
\end{figure}

In order to investigate on that point, we built a map displayed in Fig.
(5), constructed from the successive differences between 

\begin{equation}
P_k = { {\sum_{j=1}^{N_{\rm obs}}} {\sum_{l=1}^{k} (\vec{O}_j \cdot \vec{e}_l)}
  \vec{e}_l  \over {N_{\rm obs}}}  \,
\end{equation} 
and $\bar{O}$. It is quite clear that about 50 eigenprofiles should be
taken into account in order to recover, from a PCA analysis, a
pseudo-line comparable to the SLA (i.e., $\bar{O}$) or LSD ones.  This
result is in clear contradiction with the comments made in \S5.1 of
Mart\'{i}nez Gonz\'alez et al. (2008) about $P_1$ and LSD or SLA
pseudo-profiles, using noisy but synthetic data. Indeed, PCA denoising
can be made equivalent to the line addition technique, as well as to
least-squares deconvolution, but for the TBLegacy data we have been
using in that study, at the price of considering \emph{a set of
  eigenprofiles} and \emph{not} the only one associated with the
largest eigenvalue of $\vec{C}$ (and similar behaviours were noticed
for II Peg and $\varepsilon$ Eri data).

In order to understand this behaviour, it can be worthwhile analysing,
in addition to the polarization data, the so-called ``null'' spectra,
$N(\lambda)$, which comes along with standard Narval (and Espadons)
data. It is indeed customary now in stellar spectropolarimetry to
proceed with a \emph{double beam-exchange} method which consists in
recording a sequence of 4 sub-exposures associated to 2 distinct and
opposite polarization states (see e.g., Semel \& Li 1996). $N$
profiles result from a combination of sub-exposures, similar to the
one used for the extraction of the polarization signal, but on the
contrary \emph{removing} any polarization signal of astrophysical
origin. Its main usage is for the eventual detection of any spurious
signal in the data which may corrupt the astrophysical
signal. However, for clean observations (i.e., when $N(\lambda)$ is
structureless), it basically contains noise, at the same level as the
one which remains in the polarized spectra.

\begin{figure}
  \centering
  \includegraphics[width=9cm,angle=0]{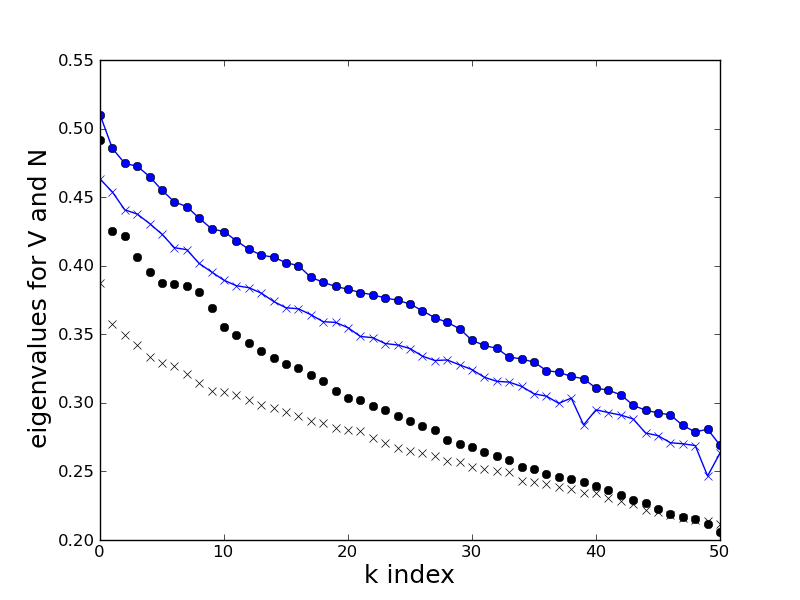}
  \caption{Successive eigenvalues of the cross-correlation matrices
    computed respectively from the $V$ (dotted lines) and the $N$
    (crossed lines) data of II Peg (discontinuous lines) and
    $\varepsilon$ Eri (continous lines) from TBLegacy. Eigenvalues for
    $\varepsilon$ Eri were magnified by a factor of 10.}
  \label{Fig6}
\end{figure}

The number of eigenprofiles to consider for the reconstructed $P_k$ to
be comparable to LSD or SLA pseudo-profiles is roughly given by the
index at which the sequences of eigenvalues of $V$ and $N$,
respectively, do overlap. Figure (6) displays two sets of eigenvalues,
which overlap indeed for $k \approx 50$ in these case. II Peg data is
represented by discontinuous lines, while $\varepsilon$ Eri data is
represented by continous lines (note also that for this latter set of
data, eigenvalues were magnified by a factor of 10). In both case,
$V$-eigenvalues correspond to dotted lines while cross symbols are for
$N$-eigenvalues. The same kind of an empirical criteria was advanced
by Mart\'{i}nez Gonz\'alez et al. (2008) during their discussion about
respective PCA analysis of a ``correlated'' (synthetic) data set and
another one of uncorrelated (Gaussian) noise.

In summary, the \emph{simultaneous} PCA analysis of $V$ and $N$ allow
for (a) the \emph{detection} of a polarized signature in the data, if the
condition

\begin{equation}
{s_{i}^{(V)}}  >  {s_{i}^{(N)}}
\end{equation}
is satisfied and (b) the \emph{characterization} of a single
representative signature, similar to LSD or SLA pseudo-profiles, which
can be made considering projections of the original data on a number
of eigenprofiles which will be given by this index at which the two
sequences of eigenvalues $s_i^{(V)}$ and $s_i^{(N)}$ do overlap.

\section{Intrinsinc dimension of the dataset}

We finally evaluate the intrinsic dimensionality of our main II Peg
and $\varepsilon$ Eri data sets, following the analysis exposed in
Asensio Ramos et al. (2007) and illustrated with synthetic data and
solar spectropolarimetric observations. To this end, we computed
maximum likelihood dimension estimators $\hat{m}$ for different values
$n$ of neighbours, for each of the profiles contained in the
observations matrix $\vec{O}$.

We adopted the formula modified by MacKay \& Ghahramani
(2005)\footnote{http://www.inference.phy.cam.ac.uk/mackay/dimension/
  -- see also Eq.\,(5) in Asensio Ramos et al. (2007)} after the
initial work of Levina \& Bickel (2005).  Both for II Peg and
$\varepsilon$ Eri data we have been analysing, values for $\hat{m}$
appear in a range of the order of 38-48, for $n$ ranging from 3 to
75. This is quite consistent with our PCA analysis of $P_{k}$
vs. $\bar{O}$ showing that our noisy data force us to consider more
eigenprofiles than a priori expected, according to Mart\'{i}nez
Gonz\'alez et al. (2008).


\section{Conclusion}

We have experimented different methods of analysis of multiline
polarized spectra of stars. We have shown, using real data, that the
simple line addition technique (Semel et al. 2009) allows for the
computation of pseudo-profiles very similar to the ones computed by
least-squares deconvolution. It is also much simpler to implement and
it requires less external input data, which makes it both simple and
efficient, and therefore very suitable for the implementation of a
standard post-processing tool for the TBLegacy database content.

From our study, LSD does not show any clear advantage on
SLA. Furthermore, its systematic use for stellar spectropolarimetric
databases would require, for the sake of interoperability, the set-up
of a specific protocol concerning the line depth cut-off criteria and
the normalization of weights used both for Stokes $I$ and $V$ data
processing.

We have also applied PCA denoising to real (and noisy) observational
data, which proves indeed very efficient. We have shown that it can
provide an alternative to SLA or LSD post-processing methods, for the
characterization of the polarization content of the multiline
observations, once the necessary number of eigenprofiles of the
cross-product matrix of the observations have been carefully
estimated. The latter can be derived from the combined PCA analysis of
$V$ and $N$ data.  Finally, and \emph{as well as for SLA}, it is in
principle equally \emph{applicable to all kind of polarization
  signals}, whatever is their physical origin or the kind of observed
state of polarization, circular or linear, which was observed.

\begin{acknowledgements}
  This research has made use of the VizieR catalogue access tool, CDS,
  Strasbourg, France. The original description of the VizieR service
  was published in A\&AS 143, 23.  Narval data were provided by the
  Cnrs/Insu Bass2000-CDAB datacenter operated by the
  \emph{Universit\'e Paul Sabatier, Toulouse}-OMP (Tarbes, France;
  {\tt http://tblegacy.bagn.obs-mip.fr/}). 
\end{acknowledgements}

\end{document}